# Suppression of spin-torque in current perpendicular to the plane spin-valves by addition of Dy cap layers

*S. Maat, N. Smith, M.J. Carey, and J.R. Childress*

*San Jose Research Center, Hitachi Global Storage Technologies,*

*San Jose, CA 95135, USA*

We demonstrate that the addition of Dy capping layers in current perpendicular to the plane giant magneto-resistive spin-valves can increase the critical current density beyond which spin-torque induced instabilities are observed by about a factor of three. Current densities as high as $5 \times 10^7$ A/cm$^2$ are measured provided that the electron current flows from the free to the reference layer. While Dy capped samples exhibit nonmagnetic 1/f noise, it is sufficiently small to be unimportant for read head operation at practical data rates.

As magnetic recording densities continue to increase and read sensor dimensions to shrink accordingly, there is an interest to replace high resistance (~ 1 Ω-μm$^2$) current perpendicular to the plane (CPP) tunnel magneto-resistive (TMR) read heads with all-metallic CPP giant magneto-resistance (GMR) read heads, which typically exhibit resistance-area (RA) products near or about 50 mΩ-μm$^2$. However, in spite of recent progress increasing the magneto-resistance values in CPP-GMR heads,[1-3] these heads also suffer from a high susceptibility to spin-torque induced instabilities.[4] While these spin-torque-induced instabilities find applications such as spin-torque magnetic random access memory and microwave-oscillators, they contribute to the noise in CPP-GMR sensors. Hence, for CPP-GMR sensors it has become an objective to enhance the critical current densities ($j_{crit}$) at which spin-torque instabilities are induced. Recent work showed that the use of antiparallel-coupled free layers[5,6] and the use of dual spin-valves[7,8] are two possibilities to enhance $j_{crit}$ and hence magnetic stability.

The use of materials with high magnetic damping offers yet another possibility: It has been shown that doping NiFe with low concentrations of rare earth elements like Dy and Ho increases its Gilbert damping parameter.[9,10] Since $j_{crit}$ is proportional to the magnetic damping parameter of the layer that is destabilized, it is desirable to increase the magnetic damping of the free layer. However, doping NiFe with elements like Dy or Ho which are non-magnetic at room temperature will dilute its magnetization and give rise to high spin-orbit coupling which will have an adverse effect on the magneto-resistance.

Thus, instead of using the rare earth as a dopant, we studied CPP spin-valves with rare earth cap layers. The structures explored are: underlayers/ IrMn/ Co$_{50}$Fe$_{50}$(25)/

Ru(7)/ Co$_{50}$Fe$_{50}$(25)/Cu(40)/ Co$_{50}$Fe$_{50}$ (7) /Ni$_{86}$Fe$_{14}$(40)/ Dy(x)/ Ru, where x was 0, 5,10, and 15 (all values are in Å). All spin-valve films were deposited by DC magnetron sputtering in an Ar-atmosphere of 2 mTorr and then annealed at 245 °C for 5 hours at 13 kOe to exchange-bias the pinned layer. The films were then patterned into pillars with diameters ranging from 40-200 nm by using a combination of electron beam lithography and ion milling. After milling, the sides of the pillars were insulated by deposition of Al-oxide, and low-resistance Au top lead contacts were made to the top of the pillars. Multiple devices of various dimensions were produced. The lead resistance and the resistance-area product, RA, were obtained from linear fits to average resistance vs. 1/(A), where A is the pillar cross section. All measured RA, and ΔR/R values are then corrected for the measured lead resistance. The Dy layers add significantly to the sensor resistance, an increase in RA of 0.38±0.02 mΩμm$^2$ / Å of Dy was observed, while ΔRA, however, remained constant at about 0.52±0.01 mΩμm$^2$ for all samples as shown in Figure 1. This behavior can be expected from the short spin-diffusion length in NiFe,[11] which is of similar magnitude as the 40 Å NiFe used in our samples. The measured increase in RA with Dy thickness is about four times larger than what can be expected from the previously reported resistivity (~94 μΩ-cm) of bulk polycrystalline Dy.[12] While thin films often exhibit a larger than bulk resistivity, the measured increase in CPP stack resistance may also dependent on chemical inter-diffusion and alloying and spin accumulation effects. For instance, the saturation magnetization of the free layer is reduced by about 0.5% per Å Dy which we attribute to inter-diffusion of Dy and NiFe. Figures 2(a) and (b) show the resistance-field (R-H) curves of spin valves acquired in magnetic fields of +/-2 kOe having (a) a Ru cap layer and (b) a Dy (15)/ Ru cap layer.

The measured devices have a 50 nm circular cross-section and were measured at $-1\cdot10^7$ A/cm$^2$, $-4\cdot10^7$ A/cm$^2$ (electron current flowing from the free to the reference layer) and $+2\cdot10^7$ A/cm$^2$ (electron current flowing from the reference to the free layer). Positive electron current destabilizes the free layer when in the AP-state, while negative electron current destabilizes the free layer in the parallel state. Negative electron current polarity results in a higher critical current density as only the electrons reflected at the reference layer are destabilizing the free layer magnetization. For both devices, when measured at low current density, the R-H curves are well behaved with the free layer switching from the low resistance parallel to the high resistance antiparallel state. At $+2\cdot10^7$ and $-4\cdot10^7$ A/cm$^2$ the R-H curve of the Ru capped spin-valve exhibits significant spin-torque induced excitations as evidenced by the field-dependent, time averaged deviations from the anti-parallel or parallel state, respectively. In contrast the spin-valve with the Dy(15)/ Ru cap is still well behaved at $+2\cdot10^7$ and $-4\cdot10^7$ A/cm$^2$ and exhibits no sign of instability, and yields about the same ΔRA as for the lower current density.

Voltage-bias dependent ΔRA values for the spin-valves under investigation were measured with the applied voltage increased in 5 mV increments using both positive and negative current polarities. After each 5 mV (-5mV) voltage increase, a magneto-resistance loop was measured at a low voltage $V_0$=5 mV (-5 mV) to make sure that only reversible spin-torque effects and no irreversible changes like physical breakdown were measured at the preceding higher voltage measurement. A change of ΔRA or RA at $V_0$=5 mV (-5 mV) after a high voltage measurement would indicate irreversible change. For the ΔRA versus current density plot in Figure 3 voltage was converted into current density, j, after correcting for lead resistance. Solid symbols in Figure 3 represent the ΔRA values measured at increasing current density (voltage), while open symbols represent the "remanant" ΔRA measured at $V_0$=5 mV (-5mV). No irreversible changes were observed for any device in the measurement range employed. At low current densities the spin-valve displays stable and well-defined parallel and antiparallel states as a function of applied fields. However, as the current density is increased past the spin-torque current density threshold, $j_{crit}$, spin-torque induced excitations result in field-dependent, time averaged deviations from the parallel and antiparallel states of the spin-valve, and the measured ΔRA decrease below the remanant ΔRA measured at -5 mV. The dashed lines indicate the bias conditions under which spin-torque induced instabilities are observed. For negative electron current polarity higher $j_{crit}$ is observed than for positive polarity. The data of Fig. 3 indicate that $j_{crit}$ is increased to about $-5 \cdot 10^7$ A/cm$^2$ in the negative electron current direction and to about $+2 \cdot 10^7$ A/cm$^2$ in the positive electron current direction by addition of a 15Å Dy damping layer. The free layer moment is slightly decreased due to the addition of the Dy cap from which a decrease in $j_{crit}$ would be expected. We attribute the enhancement of $j_{crit}$ to an increase in free layer magnetic damping due to inter-diffused Dy atoms.

Figures 4 (a) and (b) present measurements of RMS noise power (at 10 MHz) vs. continuously varied bipolar applied current (using methods described earlier[4]) for selected devices with a Ru and a Dy(15)/ Ru cap, respectively. External magnetic fields were applied at -900 Oe or -600 Oe to align the free layer magnetization parallel to that of the reference layer (P-state), and, +300 Oe or +600 Oe to align the magnetizations in the antiparallel state (AP-state). The instability is accompanied by a rapid, nonlinear increase in noise with electron current $I_e$ above the residual electronics background ($\sim 0.1\,\mathrm{nV}/\sqrt{\mathrm{Hz}}/\mathrm{mA}$). For simplicity, we will here designate $0.3\,\mathrm{nV}/\sqrt{\mathrm{Hz}}$ as the excess noise threshold above which the free-layer is considered to have passed the critical current into instability. While we observe for the two polarities critical currents of -0.27 and +0.14 mA in the spin-valve with the simple Ru cap, we observe substantially enhanced critical currents of -0.92 and +0.45 mA in the spin-valve with the Dy(15)/ Ru cap. Thus, free-layer critical currents in both P-state and AP-state are enhanced by roughly a factor of three through use of Dy cap layers. This should translate directly into a similar factor of three in output signal from the read sensor under stable operating conditions.

The unusually large apparent residual background noise observed in Fig. 4(b) led to the spectral measurements at 1 mA bias current and -900 Oe external magnetic field (P-state) shown in Fig. 5, which indicate its origin to be a 1/$f$ type resistance fluctuation noise whose rms amplitude scales linearly with bias current. Comparing the data for the spin-valves with a Ru and a Dy(15)/Ru cap clearly shows this 1/$f$ to be associated with the presence of the Dy cap. At fixed bias current, the amplitude of this 1/$f$ noise was additionally found to be independent of free-layer magnetization or applied field strength, and thus presumably nonmagnetic in character and originating at the free-layer/Dy interface. As a practical matter, however, its magnitude at $I_e = 1\,\mathrm{mA}$ ($5 \cdot 10^7\,\mathrm{A/cm}^2$) will be exceeded by ordinary electronics noise above 10 MHz, and for most if not all anticipated recording system data rates should not compromise the broadband advantage in signal output afforded by the three-fold increase in critical current using a Dy cap layer.

In conclusion, we demonstrate that the addition of Dy capping layers in current perpendicular to the plane giant magnetoresistive spin-valves present a viable option for increasing, by about a factor of three, the critical current density $j_{crit}$ beyond which spin-torque induced instabilities are observed. While Dy capped samples exhibit low frequency $1/f$ noise, this source of excess noise should not be significant in most present and future recording system applications where data rates are at least several hundred MHz.

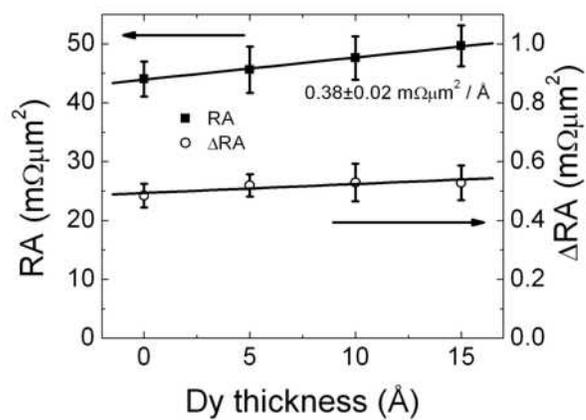

**Figure 1** RA and ΔRA versus Dy layer thickness. While ΔRA remains constant at about 0.52±0.01 mΩμm$^2$, an increase of 0.38±0.02 mΩμm$^2$ / Å Dy was observed for RA.

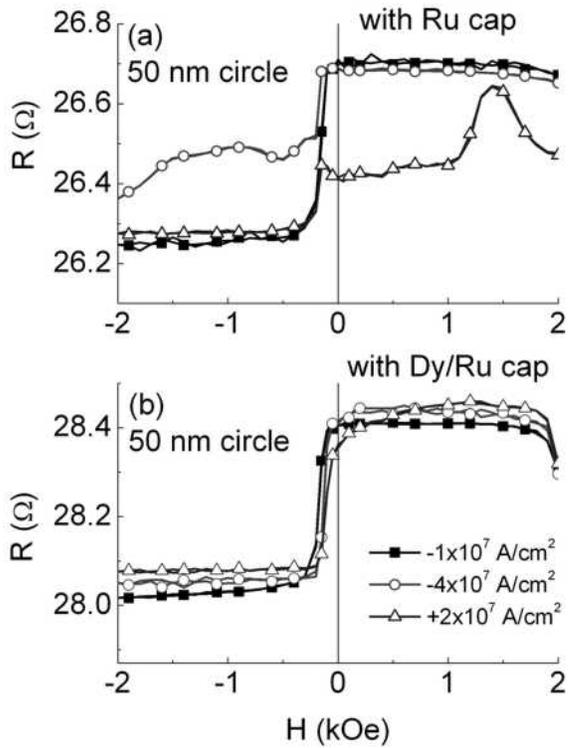

**Figure 2** Resistance field (R-H) curves for a spin-valve with (a) a Ru and (b) a Dy(15)/Ru cap at a current density of $-1 \cdot 10^7$ A/cm$^2$, $-4 \cdot 10^7$ A/cm$^2$ (Current is flowing from the free to the reference layer) and $+2 \cdot 10^7$ A/cm$^2$ (Current is flowing from the reference to the free layer). While the spin-valve without the Dy layer exhibits spin-torque induced instability at $-4 \cdot 10^7$ A/cm$^2$ and $+2 \cdot 10^7$ A/cm$^2$ the spin-valve with the Dy cap layer exhibits a well behaved R-H curve at the same current densities.

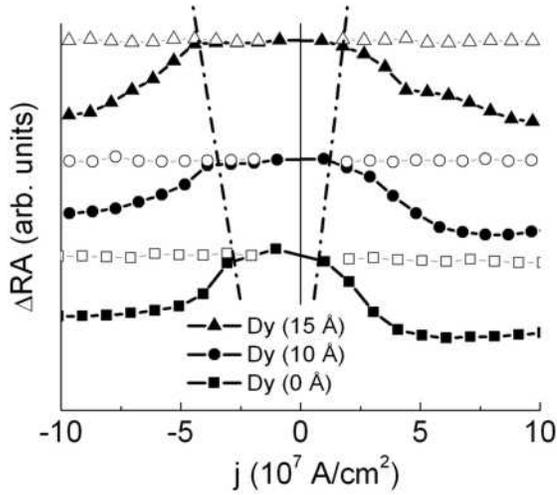

**Figure 3** ΔRA vs. current density for the spin-valves with a Ru/Dy(x), x=0, 10, 15 Å caps. The critical current density is increased by addition of a Dy layer. Solid symbols represent ΔRA measured at bias voltage, while open symbols represent ΔRA measured at low bias voltage after removing the high bias. The ΔRA data is offset for clarity. The dashed lines indicate the bias conditions under which spin-torque induced instabilities are observed.

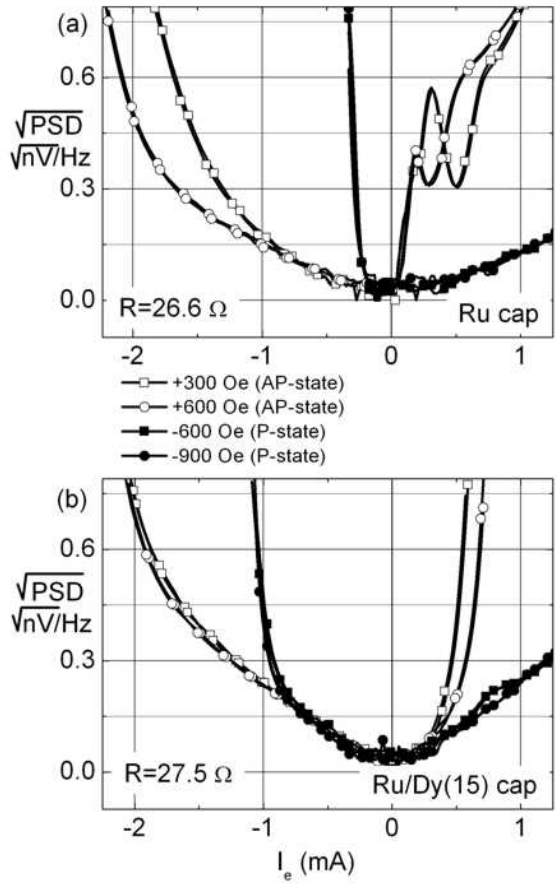

**Figure 4** RMS noise power spectral density (PSD) versus electron current $I_e$ for spin-valves with (a) a Ru and (b) a Ru/ Dy(15) cap, measured at external applied fields of -900 Oe or -600 Oe (P-state),and +300 Oe or +600 Oe (AP-state). The background noise measured at zero electron current has been subtracted.

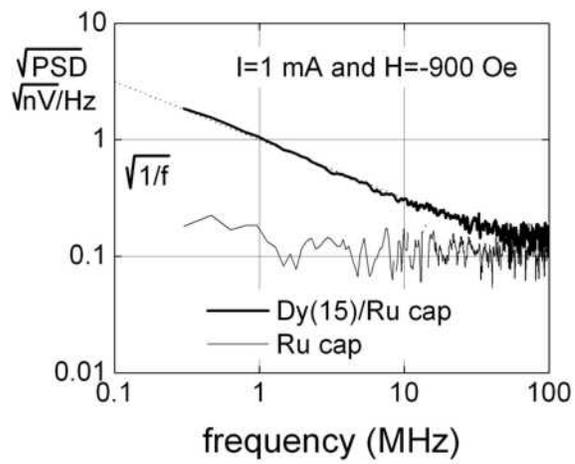

**Figure 5** Noise PSD for spin-valves with a Ru and a Dy(15)/Ru cap. The spin-valve with the Dy layer exhibits 1/*f* noise.